\documentstyle[12pt]{article}
\input math_macros
\newcommand {\geapprox}{\;_{\sim}^{>}\;}
\newcommand {\leapprox}{\;_{\sim}^{<}\;}

\begin{document}

\begin{titlepage}
\begin{center}
\today      \hfill      LBL-35103 \\
            \hfill      UCB-PTH-94/04 \\
\vskip .5in

{\LARGE  The Constituent Quark as a \\ Topological Soliton}
\footnote{This work was supported in part by the Director, Office of
Energy Research, Office of High Energy and Nuclear Physics, Division
of High Energy Physics of the U.S. Department of Energy under Contract
DE-AC03-76SF00098 and in part by the National Science Foundation under
grant PHY-90-21139.}

\vskip .3in

{\large Gregory L. Keaton}

\vskip .3in

{\it Department of Physics, University of California, Berkeley \\
and \\
Theoretical Physics Group, Lawrence Berkeley Laboratory \\
Berkeley, CA 94720}
\end{center}

\vskip .5in

\begin{abstract}
Recently it was proposed that the constituent quark is a topological soliton.
I
investigate this soliton, calculating its mass, radius, magnetic moment, color
magnetic moment, and spin structure function.  Within the approximations used,
the magnetic moments and spin structure function cannot
simultaneously be made to agree
with the constituent quark model.  Some discussion of what to expect from
better
approximations is included.
\end{abstract}

\end{titlepage}

\newpage
\renewcommand{\thepage}{\arabic{page}}
\setcounter{page}{1}

\section{Introduction}

Why does the constituent quark model work?  This paper examines one possible
answer to that question.

The constituent quark model is puzzling because all the clues from QCD and
current algebra \cite{bluegeorgi}
indicate that the nucleon is made of a swarm of gluons and
nearly massless quarks and anti-quarks, all interacting strongly.  According
to the constituent quark model, however, the nucleon is composed of nothing but
three massive quarks, interacting weakly.  The model works well:  baryon
\cite{glashow,quigg} and
heavy meson \cite{lebed}
masses can be fitted to within a few percent, and most baryon magnetic
moments \cite{bluegeorgi,quigg} to within 30\%.

One explanation for the success of this model is that the nucleon really does
contain three weakly interacting components.  In this picture, the nearly
massless ``current quarks'' are fundamental particles, and through their strong
interactions each is able to draw around itself a cloak of virtual
gluons and quark--anti-quark pairs, resulting in the collective excitation
called a ``constituent quark''.  The constituent quark has the same spin and
flavor as the original current quark, but is heavier and less strongly
interacting.

To explain how the constituent quark might arise as a collective excitation,
two
models have been proposed.  The first is Manohar and Georgi's chiral quark
model
\cite{georgimanohar} (which is closely related to the Nambu--Jona-Lasinio model
\cite{njl}).  In this model the current
quark increases its mass by coupling to the quark condensate that forms when
chiral symmetry is broken.  The resulting constituent quark then automatically
has the same spin and flavor as the original current quark.

The second model is the quark soliton model proposed by Kaplan \cite{kaplan}.
It is based on a very simple and
appealing idea:  the quark condensate may undergo rotations in color space as
well as flavor space.  The flavor rotations give rise to the usual non-linear
sigma model Lagrangian used in pion physics.  One form of this Lagrangian,
the Skyrme Lagrangian \cite{skyrme,anw,balachandran},
will support a topological soliton which is a model of
the nucleon.  Similarly, the color rotations of the condensate can be described
by a Lagrangian which also supports a soliton.  This soliton is a candidate for
the constituent quark.
The topological properties of this winding number 1 soliton ensure that it has
spin 1/2 and baryon number 1/3, just as the original current quark.

This soliton has been analyzed in two dimensions \cite{2d}.
In four dimensions,
its mass and radius have been computed \cite{gks},
and it was found that either the
soliton's mass or its radius (or both) must be larger than expected for the
constituent quark.  However, Ref. \cite{gks} argues that this is not a fatal
flaw in the model.

In this paper I have used a different technique to study the mass and radius,
and have reached similar conclusions.  I have gone on to evaluate the soliton's
magnetic moment, color magnetic moment, and spin structure function.
Within the approximations used, the spin structure function and the magnetic
moments cannot both be fitted simultaneously
in this model.  Some speculation is offered
on what to expect from better approximations.

The main purpose of this paper is to show how the
static properties of the soliton may be evaluated, and to test whether the
soliton's properties are compatible with the constituent quark's.
Section \ref{soliton} introduces the
Lagrangian and the soliton.  Section \ref{properties} shows how I extract the
static properties of the soliton.  These properties are then computed and the
results given in Section \ref{results}.
The paper closes with a short discussion
in Section \ref{discussion}.

\section{The Soliton}
\label{soliton}

In the quark soliton model,
the quark condensate can undergo rotations in color space as well
as in flavor space. The color degrees of freedom
are parametrized by $U = e^{2 i
\Pi^{a} T^{a}/f}$.  The capital $\Pi^{a}$ is used to distinguish this field
from
the ordinary pion field $\pi^{a}$; the $T^{a}$ are the generators of color
SU(3); the constant $f$ is analogous to the pion decay constant $f_{\pi}$.
With $\tilde{R_{\mu}} \equiv U^{\dag} D_{\mu} U$, Kaplan \cite{kaplan} proposed
the following one-flavor Lagrangian
(which could be extended to more flavors later) :
\begin{eqnarray}
{\cal L}[U,A_{\mu}]  & = & - \frac{1}{4 g^{2}} G^{\mu \nu}_{a} G^{a}_{\mu \nu}
- \frac{f^{2}}{4} tr (\tilde{R_{\mu}} \tilde{R^{\mu}}) +
\frac{1}{32 e^{2}} tr ( [\tilde{R_{\mu}},\tilde{R_{\nu}}]^{2}) \nonumber \\
 &  &   + \frac{2}{3}f^{4}\nu^{2}tr(T_{a}UT_{a}U^{\dag}) + n{\cal L}_{WZ}
\label{l1}
\end{eqnarray}
This Lagrangian is patterned after the Skyrme Lagrangian
\cite{skyrme,anw,balachandran}.
The first term gives kinetic energy to the gluons; the second gives kinetic
energy to the chiral field U.  The third term, called the ``Skyrme term'', is
introduced to stabilize the soliton solution.  If this term is absent from the
ordinary (ungauged) Skyrme Lagrangian, then the soliton shrinks to zero
size.  The fourth term breaks color $SU(3)_{L}\times SU(3)_{R}$ symmetry, and
consequently gives mass to the (as yet undiscovered) $\Pi$ particles.  This
term
is necessary because QCD interactions, whose low energy behaviour this
Lagrangian is intended to model, explicitly break {\em color} (not flavor!)
$SU(3)_{L}\times SU(3)_{R}$ symmetry.  The last term of Eq.(\ref{l1}) is the
Wess--Zumino term \cite{witten,pakrossi}, which
takes anomalies into account.

Each of the first four terms is multiplied by an arbitrary constant, to be
fitted
phenomenologically.  $g,\:e,\:$ and $\nu$ are dimensionless; $f$ has the
dimensions of mass.
The mass of
the $\Pi$ particle is equal to $2 \nu f$.  The integer coefficient of the
Wess--Zumino--Witten anomaly term
is $n=1$, as opposed to $n=N_{c}=3$ in the
Skyrme Lagrangian.
The soliton solutions of this Lagrangian
have baryon number $n/N_{c}$ \cite{kaplan}, and upon rotation by $2 \pi$,
the soliton will acquire a phase
$(-1)^{n}$ \cite{witten}.  Therefore the soliton
is a fermion with baryon number 1/3.
This soliton, which Kaplan calls a ``qualiton'', is a candidate for the
constituent quark.

The construction of the soliton from the Lagrangian (\ref{l1}) proceeds in
three
steps which will be sketched below (see \cite{kaplan} for more details).
Step one:  construct the classical solution.  The field U takes on the
``hedgehog'' ansatz form:
\begin{equation}
U_{cl} = e^{i F(r) {\bf \hat{r}} \cdot \mbox{\boldmath $\tau$}  }
\label{ucl}
\end{equation}
where the $\tau^{i}$ are the Pauli matrices embedded in color SU(3).
In this ansatz $F(0) = \pi$ and $F(\infty) = 0$.
The gauge field is given by
\begin{equation}
A_{i\:cl} = - A^{i}_{cl} = i \frac{\gamma (r)}{2 r} \epsilon_{ijk}
\hat{r}^{j} \tau^{k} \;\;,\;\;\;\;    i=1,2,3
\end{equation}
Throughout this paper an anti-hermitian gauge field is used:  $D_{\mu} =
\partial_{\mu} + A_{\mu}$.
The profile functions $F$ and $\gamma$ can be determined by minimizing the
soliton's classical mass, $m_{cl} = -{\cal L}[U_{cl},A_{i\:cl}]$.  The
resulting
Euler--Lagrange equations can be solved numerically \cite{colsys}.  It is
convenient to use the dimensionless variable $\tilde{r}=2fr$, since the
Euler-Lagrange equations that determine $F(\tilde{r})$ and $\gamma(\tilde{r})$
do not depend on the parameter $f$.  Fig. 1 shows the profiles $F(\tilde{r})$
and $\gamma(\tilde{r})$ for $1/e = 0,\: g = 12.4,\:$ and $\nu = 237$.  These
values were chosen for two reasons:  first, the resulting soliton has the same
spin structure function as expected for the constituent quark (see below).
Second, this demonstrates that the soliton can be stable even when the Skyrme
term is absent ($1/e = 0$).  Evidently, the gauge field is sufficient to
stabilize the soliton.  (A similar feature has been seen
in the Skyrme Lagrangian, where
the Skyrmion solution itself is stable in the absence of
the Skyrme term as long as the $\rho$-meson gauge field is present\cite{rho}.)

Step two of constructing the soliton:  make it rotate.  This is done by
conjugating the field $U_{cl}$ by a matrix $\Omega$:
\begin{equation}
U=\Omega U_{cl} \Omega^{\dag}
\end{equation}
If the gauge field were absent, $\Omega$ would be a function of time but not of
space.  Since the gauge field is present, however, $\Omega$ must depend on $r$
as well as $t$.  This can be understood as follows:  the gauge field rotates
also:
\begin{equation}
{\bf A} = \Omega {\bf A}_{cl} \Omega^{\dag} - \Omega \nabla \Omega^{\dag}
\label{rota}
\end{equation}
The rotation of $U$ and ${\bf A}$ generates a charge density
\begin{equation}
j_{0}^{a} = i \frac{f^{2}}{2} tr[ (U^{\dag} T^{a} U - T^{a}) \tilde{R}_{0} ]
- \frac{i}{8e^{2}} tr \{ [ (U^{\dag} T^{a} U - T^{a}),\tilde{R}^{\nu}]
[\tilde{R}_{0},\tilde{R}_{\nu}] \}
\label{j}
\end{equation}
The color electric fields must be given by the rotating gauge fields of Eq.
(\ref{rota}), and must also satisfy Gauss's Law with the charge of Eq.
(\ref{j}).  This constrains $\Omega$ to satisfy certain differential equations
given in Ref. \cite{kaplan}.  For now it is enough to know that $\Omega$ can be
parametrized by three functions $\omega_{1}(r),\:\omega_{2}(r),\:$ and
$\omega_{3}(r)$.  These functions are shown in Fig. 2 for the same set of input
parameters as in Fig. 1. $\omega_{1}$ and $\omega_{3}$ are closely analogous to
ordinary angular velocity, and they are smaller for smaller $r$.  That is, the
soliton must rotate more slowly in the middle than on the outside because
otherwise it generates too much charge to be consistent with Gauss's Law.

As $r \rightarrow \infty$, the matrix $\Omega(r,t)$ becomes equal to a matrix
$W(t)$.  The Lagrangian (\ref{l1}) can be rewritten in terms of this matrix as
\cite{kaplan}:
\begin{equation}
L = -m_{cl} + \frac{I_{1}}{2} \sum_{m=1,2,3}(i W^{\dag}\dot{W})^{2}_{m} +
\frac{I_2}{2} \sum_{\alpha=4 \ldots 7} (i W^{\dag} \dot{W})^{2}_{\alpha} +
\frac{1}{\sqrt{12}}(i W^{\dag} \dot{W})_{8}
\label{ltop}
\end{equation}
Here $(W^{\dag} \dot{W})_{a} \equiv 2 \: tr ( T_{a} W^{\dag} \dot{W})$.
The last term in this Lagrangian comes from the Wess--Zumino term.
The moments of inertia $I_{1}$ and $I_{2}$ are coefficients that can be
computed once $F,\:\gamma,\:\omega_{1},\:\omega_{2},\:$ and $\omega_{3}$ are
known.

The Lagrangian (\ref{ltop}) describes a top spinning in SU(3) space.  When the
Hamiltonian is constructed, the canonical momenta $P_{a}$ are used.
Explicitly \cite{balachandran}, $P_{a} = I_{1} (i W^{\dag} \dot{W})_{a} $ for
$a = 1,2,3\:$ and $P_{a} = I_{2} (i W^{\dag} \dot{W})_{a} $ for $a = 4,\ldots,
7$.  $P_{a}$ is equal to the ordinary angular momentum $\Lambda_{a}$ for $a =
1,
2,3$.
Since the Wess--Zumino term only contributes one power of $(W^{\dag} \dot{W})
_{8}$ to the Lagrangian (\ref{ltop}), $P_{8}$ will not have an equation of
motion but rather an equation of constraint:  $P_{8} = 1/\sqrt{12}$.

Step three in building the soliton:  quantize it.  Here $W$ and $P_{a}$ are
no longer treated
as ordinary matrices, but as operators on a Hilbert space.
The total mass of the soliton is given by the resulting
energy eigenvalues,
\begin{equation}
E = M_{TOT} = m_{cl} + j(j+1)(\frac{1}{2I_{1}} - \frac{1}{2I_{2}})
 + \frac{1}{2 I_{2}} (C_{2}-\frac{1}{12})
\label{mtot}
\end{equation}
where $j$ is the spin quantum number and $C_{2}$ is the color SU(3) casimir.
Since we are interested in a spin--1/2, color triplet particle, $j=1/2$ and
$C_{2} = 4/3$.  Then
\begin{equation}
M_{TOT} = m_{cl} + \frac{3}{8 I_{1}} + \frac{1}{4 I_{2}}
\label{mtot2}
\end{equation}

The soliton is described by a state $|q,\sigma>$ whose wave functions are given
by the SU(3) Wigner D-functions in the triplet representation
\cite{balachandran,adkinsnappi,guadagini}:
\begin{equation}
\psi_{q,\sigma}(W) = <W|q,\sigma> = \sqrt{3} D^{(3)}_{q,\sigma} (W)
\label{psi}
\end{equation}
Here $q$ and $\sigma$ are SU(3) indices:  $q=(I,I_{3},Y)$ gives the color
isospin and hypercharge of the particle, and $\sigma = (s,-m_{s},1/3)$ gives
the
spin.  The last entry is constrained to be 1/3 by the Wess--Zumino term.

There is one point worth mentioning now.  In the above procedure, the functions
$F$ and $\gamma$ are determined by minimizing the classical mass; they are then
used to calculate the moments of inertia $I_{1}$ and $I_{2}$.  This is called
the ``semi-classical'' approach.  A more exact procedure \cite{braaten} is to
view the total mass as a functional of $F$ and $\gamma$,
\begin{equation}
M_{TOT}[F,\gamma] = m_{cl}[F,\gamma] + \frac{3}{8 I_{1}[F,\gamma]} +
\frac{1}{4 I_{2}[F,\gamma]}
\label{totquant}
\end{equation}
and then find those functions $F$ and $\gamma$ which minimize the total mass,
not just the classical mass.  The resulting integro-differential equations have
never been worked out.

In the Skyrme model the semi-classical approach is sufficient because $m_{cl}$
is of order $N_{c}$ and the moments of inertia are also of order $N_{c}$.
Therefore the rotational energy does not contribute much to the total energy,
and the error made in the semi-classical approximation is small.  However, no
such $N_{c}$--counting argument exists for the qualiton, and there is no
guarantee that the semi-classical approach is enough.  Still, it is worth
trying.

Having constructed the soliton, then, the question is
whether the four parameters $f,\:e,\:g,\:$ and $\nu$ can be adjusted to give
realistic values for the static properties of the constituent up and down
quarks.

\section{Properties of the Soliton}
\label{properties}

The static properties of the soliton are discussed in this section.  For each
observable, I first state what is expected from the static quark model, and
then describe how to extract this quantity from the soliton model.  The results
are then used in Section \ref{results} for numerical
computations.
\\[1 cm]
{\bf Mass and Radius}
\\*

The constituent quark mass is typically taken to be $\approx 350\:$ MeV.  For
definiteness, I will take Quigg's value, $m = 362\:$ MeV \cite{quigg}.  The
radius
of the constituent quark should be less than the radius of the nucleon.  The
isoscalar rms radius of the nucleon is $r_{rms} = 0.72\:$ fm.
Therefore, in units where $\hbar = c = 1$, there is an upper bound on the
dimensionless quantity $m \cdot r_{rms}$ for the constituent quark:
\begin{equation}
m r_{rms} \leq 1.3
\label{bound}
\end{equation}

In the soliton model, $r_{rms}$ can be computed using the singlet part of the
anomalous Wess--Zumino current:
\[
r^{2}_{rms} = \int d^{3}r \: r^{2} \: J^{0}_{WZ}
\]
where \cite{pakrossi}
\[
J^{\mu}_{WZ} = \frac{1}{48\pi^{2}} \epsilon^{\mu\alpha\beta\gamma}
tr [2 \tilde{R_{\alpha}}\tilde{R_{\beta}}\tilde{R_{\gamma}} -
3 F_{\alpha\beta}(\tilde{R_{\gamma}}+\tilde{L_{\gamma}})]
\]
Here
$\tilde{R_{\mu}} = U^{\dag} D_{\mu} U$ and $ \tilde{L_{\mu}} = (D_{\mu} U)
U^{\dag}$.  The convention $\epsilon_{0123} = -\epsilon^{0123} = 1$ is used.
Using Eqns. (\ref{ucl}) - (\ref{rota}) (cf. \cite{kaplan}),
\begin{equation}
r^{2}_{rms} = \frac{1}{\pi} \frac{1}{(2ef)^{2}} \int \tilde{r} d\tilde{r}
[2F - (1+\gamma)^{2} \sin 2F]
\label{rms}
\end{equation}
Given the radius from Eq.(\ref{rms}) and the mass as calculated from Eq.
(\ref{mtot2}), can the qualiton satisfy the inequality
(\ref{bound})?

To answer this question, it is easiest to look first in the limit where
$g \rightarrow 0$ and $\nu \rightarrow 0$, because the result is identical to
the ordinary
SU(3) Skyrmion with $N_{c}=1$. In this model the total mass is
\begin{eqnarray*}
M & = & m_{cl} + \frac{3}{8 I_{1}} + \frac{1}{4 I_{2}} \\
  & = & \hat{m} \frac{2f}{e} + \frac{3}{8} \left( \frac{2fe^{3}}{\hat{I}_{1}}
\right) +
        \frac{1}{4} \left( \frac{2fe^{3}}{\hat{I}_{2}} \right)
\end{eqnarray*}
where $\hat{m},\:\hat{I}_{1},$ and $\hat{I_{2}}$ can be calculated numerically:
$\hat{m} = 36.5,\; \hat{I_{1}}=106.6,\; \hat{I_{2}}=40.6$. (Ref.
\cite{balachandran} gets similar values.)  The isoscalar rms radius is
\[
r_{rms} = \frac{\hat{r}_{rms}}{2fe}
\]
with $\hat{r}_{rms} = 2.12$ (cf. \cite{anw}).  Therefore,
\[
M r_{rms} = \hat{r}_{rms}[\frac{\hat{m}}{e^{2}} + (\frac{3}{8 \hat{I_{1}}} +
\frac{1}{4 \hat{I_{2}}})e^{2}]
\]
Differentiating the above equation with respect to $e$ and setting it equal to
zero reveals that, when $g=\nu=0$,
\begin{equation}
M r_{rms} \geq 2.52
\label{bigmass}
\end{equation}
with the minimum occurring at e=7.84.  Numerically, it is found that the full
soliton also has a minimum $(M r_{rms})=2.52$ at this ``Skyrme'' configuration
($e=7.84,\;\nu=0,$ and $g=0$);
$(M r_{rms})$ only increases when $g$ and $\nu$ move away from 0.

Therefore it is not possible for the soliton to satisfy the inequality
(\ref{bound}).  Ref. \cite{gks} uses a different technique but comes to the
same
conclusion:  the radius or the mass of the soliton is larger than expected for
a constituent quark.  Ref. \cite{gks}
explores whether a large radius is a serious flaw. The
problem with this approach is that it requires the confining force to
be so strong that it contracts the quark to roughly half its original size.
This
goes against the spirit of the constituent quark model, in which the quarks are
weakly interacting inside the hadron.

The alternate possibility is that the mass is large.
At first it might seem that
this, too, would violate the principles of the constituent quark model, since
the binding energy per quark would be at least half the constituent mass.  This
would require the quarks to be very strongly interacting as well.  However, the
concept of binding energy does not exist in confining theories.  In fact, until
details of the inter--quark forces are included, the relationship between the
mass of the constituents and the mass of the nucleon cannot be determined.
Since these details have not yet been worked out, the question of the excessive
mass cannot be addressed in this paper.
\\[1 cm]
{\bf Magnetic Moment}
\\*

The magnetic moment of a particle with charge $q$ and spin $\bf S$ can be
written
\begin{equation}
\mbox{\boldmath $\mu$} = q \beta {\bf S}
\end{equation}
where $\beta$ is a parameter with the dimensions of length.  In the constituent
quark model (where the quarks have a Dirac
g--factor of two), $\beta = 1/m$.  Using $m = 362\:$ MeV, the value of $\beta$
is $0.544\:$ fm.

In the soliton model, the magnetic moment is given by
\begin{equation}
\mbox{\boldmath $\mu$} = q \frac{1}{2} \int d^{3}x \:{\bf r \times J}_{WZ}
\end{equation}
It is easiest to work in the gauge where ${\bf A}_{i} = {\bf A}_{i\:cl},\:
A_{0}=\Omega^{\dag}\dot{\Omega},$ and $U=U_{cl}$.  Then,
\begin{equation}
J^{i}_{WZ} = - \frac{1}{8 \pi^{2}} \epsilon^{ijk} \partial_{k}
tr [A_{0}U^{\dag}\partial_{j}U + A_{0}\partial_{j}U U^{\dag} +
A_{0}U^{\dag}A_{j}U - A_{0}UA_{j}U^{\dag}]
\label{vectjwz}
\end{equation}
$A_{0}$ can be written in terms of the angular momentum ${\bf \Lambda}$:
\begin{eqnarray*}
A_{0} = \Omega^{\dag}\dot{\Omega}&=&\frac{-i}{2I_{1}}
[(\omega_{1}+\omega_{2}){\bf \Lambda} \cdot \mbox{\boldmath $\tau$} -
\omega_{2}{\bf( \hat{r} \cdot \Lambda)\hat{r}}\cdot \mbox{\boldmath $\tau$}] \\
     & & + \mbox{ terms} \propto \lambda_{4} \ldots \lambda_{8}
\end{eqnarray*}
The extra terms proportional to $\lambda_{4} \ldots \lambda_{8}$ do not
contribute to the trace in Eq. (\ref{vectjwz}).  The resulting magnetic moment
is
\begin{eqnarray}
\mbox{\boldmath $\mu$} & = & \lim_{R \rightarrow \infty} - \frac{q}{3 \pi}
\frac{{\bf \Lambda}}{I_{1}} \left\{ \int_{0}^{R} r^2 dr [\omega_{1}F^{\prime} +
\frac{1}{r} (\omega_{1}+\omega_{2})(1+\gamma)\sin 2F] \right. \nonumber \\
  &  &  \left. +
[r^{3}\omega_{1}F^{\prime} + \frac{1}{2}r^{2}(\omega_{1}+\omega_{2})(1+\gamma)
\sin 2F]_{r=R} \right\}
\end{eqnarray}
In Section \ref{results} this formula is used for the numerical computation
of $\beta$ for the soliton.
\\[1 cm]
{\bf Color Magnetic Moment}
\\*

In the constituent quark model, the hyperfine mass splitting of the hadrons
is given by the
interaction of the color magnetic moments of the constituents:
\[
\Delta E_{hfs} = -\frac{2}{3} |\psi(0)|^{2} \sum_{i<j}
<n| \mbox{\boldmath $\mu$}^{a}_{(i)} \cdot \mbox{\boldmath $\mu$}^{a}_{(j)} |n>
\]
where the sum is over the quarks $i$ and $j$ in the nucleon $|n>$.
The color magnetic moment can be defined by
\begin{equation}
\mbox{\boldmath $\mu$}^{a} = \mu_{c} {\bf S} \lambda^{a}
\label{mudef}
\end{equation}
where the $\lambda^{a}$ are the Gell-Mann matrices and
$\mu_{c}$ is a parameter with the dimensions of length.  In the constituent
quark model, $\mu_{c} = g/2m$.
Using $\alpha_{s}=g^{2}/4\pi=0.4$ and $m=362\:$ MeV \cite{quigg},
the value of $\mu_{c}$ is $0.610\:$ fm.  The ratio of the
constituent quark's color magnetic moment to its magnetic moment is
proportional to $\mu_{c}/\beta = 1.12$.

The qualiton also has a color magnetic moment, which can be extracted from the
asymptotic behaviour of its B-field.  The standard dipole
form for ${\bf B}$, using the normalization of Eq. (\ref{l1}), is
\begin{equation}
B^{i}_{a} = \frac{g \mu^{j}_{a}}{4 \pi} (3 \hat{r}_{i} \hat{r}_{j} -
\delta_{ij}) \frac{1}{r^{3}}
\label{std}
\end{equation}
The B-field of the qualiton will turn out to have a similar form at large $r$,
so the coefficient $\mu^{j}_{a}$ can easily be determined.

At large radius, the B-field of the classical soliton is
\begin{equation}
B^{i}_{a\:cl} = -r_{B}(3 \hat{r}_{i} \hat{r}_{a} - \delta_{ia}) \frac{1}{r^{3}}
\label{bcl}
\end{equation}
The constant $r_{B}$ is determined by the numerical solution for
$\gamma:\;$ $ \lim_{r \rightarrow \infty} \gamma(r)=-r_{B}/r$.
The B-field of the quantized soliton can now be calculated.

Under the quantization procedure, ${\bf B}$ becomes an operator $\hat{\bf B}$
rather than just a matrix.  At large radius, $\hat{{\bf B}} = W {\bf B}_{cl}
W^{\dag}$.  The
expectation value of
$\hat{{\bf B}}$ with respect to the quark soliton state $|q,\sigma>$ is
\[
{\bf B}  =  <q,\sigma|\hat{{\bf B}}|q,\sigma>
         =  <q,\sigma|W{\bf B}_{cl} W^{\dag}|q,\sigma>
\]
Using $B^{i} = B^{i}_{m} T^{m}$,
\begin{equation}
B^{i}_{n} = 2<q,\sigma|tr[T^{n}WT^{m}W^{\dag}]|q,\sigma>B^{i}_{m\:cl}
\label{b}
\end{equation}
In order to compute the above matrix element, we can use the wavefunctions
given
in Section \ref{soliton}, Clebsch--Gordan coefficients \cite{clebsch}, and the
identity \cite{guadagini}
\[
<W|tr[T_{n}WT_{m}W^{\dag}]|W> = \frac{1}{2} D^{(8)}_{nm}(W)
\]
The result is
\begin{equation}
<q,\sigma|Tr[T^{n}WT^{m}W^{\dag}|q,\sigma> = - \frac{3}{32}<q,\sigma|
\sigma^{m} \lambda_{n} |q,\sigma>
\end{equation}
Combining this equation with (\ref{b}) and (\ref{bcl}) gives
\begin{equation}
B^{i}_{n} = \frac{3 r_{B}}{16} <q,\sigma|\sigma^{m} \lambda_{n}|q,\sigma>
(3 \hat{r}_{i} \hat{r}_{m} - \delta_{im}) \frac{1}{r^{3}}
\end{equation}
Therefore, using Eq. (\ref{std}),
\begin{equation}
\mbox{\boldmath $\mu$}_{a} = \frac{4 \pi}{g} \frac{3 r_{B}}{16}
\mbox{\boldmath $\sigma$}\lambda_{a}
\label{mu}
\end{equation}
Using ${\bf S} = \mbox{\boldmath $\sigma$}/2$ and Eq. (\ref{mudef}),
we find that $\mu_{c} = (4 \pi/g)(3 r_{B}/8)$.
This information is used in Section \ref{results}.

The above procedure is sufficient to determine the color magnetic moment, but
there is another way to evaluate it which parallels the calculation of the
ordinary magnetic moment.  Testing whether these two methods agree serves as a
useful check on the numerical computations.

The color magnetic moment should be given by
\[
\mbox{\boldmath $\mu$}_{a} = \frac{g}{2} \int d^{3}r\: {\bf r \times J}_{a}
\]
where $J^{\mu}_{a}$ is the current which couples to the gauge field
$A^{\mu}_{a}$.  This integral has already been worked out in the SU(2) case
\cite{anw} for $\nu,\: \alpha_{s} \rightarrow 0$.  It is unchanged for SU(3),
and
the result is
\begin{equation}
\mu^{i}_{a} = - g I_{1} tr[T^{a} W T^{i} W^{\dagger}]
\label{gumbda}
\end{equation}
where the trace is to be taken as a matrix element between quark states, as
above.  Combining this with Eq. (\ref{mu}) gives
\[
r_{B} = \frac{\alpha_{s} I_1}{2} \;\;\;\;(\alpha_{s}, \nu \ll 1)
\]
This expression is indeed satisfied by the numerical computations of $r_{B}$
and
$I_{1}$ when $\alpha_{s}$ and $\nu$ are small.
\\[1 cm]
{\bf Spin Structure}
\\*

The spin structure function of the constituent quark is not as well established
as the previous properties.  In fact, it is not obvious from the recent
spin structure experiments whether the data are even consistent with the
constituent quark model at all.  They turn out to be consistent, but some
explanation is required.

The Fourier transform of the constituent quark spin structure function is
defined as follows:
for a single spin-up quark $|q\uparrow>$ and the field $\psi$ which annihilates
it,
\[
g_{q}({\bf r}) = <q\uparrow|\bar{\psi}({\bf r})\gamma_{z}\gamma_{5}\psi({\bf
r})|q\uparrow>
\]
I will call the integral over all space of $g_{q}$ the ``spin content''
$s_{q}$:
\[
s_{q} = \int d^{3}r\:g_{q}({\bf r})
\]
In the non-relativistic limit one would expect $s_{q} = \sigma_{z} = 1$.  As we
will see below, the recently measured spin structure functions of the neutron
and proton will force us to change these expectations.

To begin, look at the nucleon.
The contribution of the up
quark spin to a spin--up proton is:
\begin{equation}
\Delta u = \int d^{3}r\:<p\uparrow|\bar{u}\gamma_{z}\gamma_{5}u|p\uparrow>
\label{delu}
\end{equation}
If the up quark is non--relativistic, $\Delta u$ is just equal to the number of
up quarks with spin parallel to the proton's spin, minus the number of up
quarks
with spin anti-parallel.  $\Delta d$ and $\Delta s$ are similarly defined.
Several relations exist between these quantities:
\begin{equation}
\Delta u - \Delta d = g_{A} = 1.2573 \pm .0028
\label{bj}
\end{equation}
is used in the Bjorken sum rule \cite{ga}, and
\begin{equation}
\Delta s - \Delta d = D - F = .328 \pm .019
\label{hyp}
\end{equation}
results from an analysis of semileptonic hyperon decay \cite{hyperon}.
Combining both of the
above two equations with a third equation would give three equations and three
unknowns, and so $\Delta u,\: \Delta d$, and $\Delta s$ could all be
determined.
Actually, any one of several equations could be chosen as the third equation in
this procedure.  The EMC experiment \cite{emc} gives
\begin{equation}
\frac{1}{2} \left( \frac{4}{9} \Delta u + \frac{1}{9} \Delta d + \frac{1}{9}
\Delta s \right) = 0.126 \pm 0.02
\label{emceq}
\end{equation}
The result of the E142 experiment \cite{e142} is
\begin{equation}
\frac{1}{2} \left( \frac{1}{9} \Delta u + \frac{4}{9} \Delta d + \frac{1}{9}
\Delta s \right) = -0.022 \pm 0.011
\label{e142eq}
\end{equation}
and the SMC experiment \cite{smc} gives
\begin{equation}
\frac{1}{4} \left( \frac{5}{9} \Delta u + \frac{5}{9} \Delta d + \frac{2}{9}
\Delta s \right) = 0.023 \pm 0.025
\label{smceq}
\end{equation}
Equations (\ref{bj}) and (\ref{hyp}) can be combined with either Eq.
(\ref{emceq}), (\ref{e142eq}), or (\ref{smceq}).  The results of all three
possibilities are given in the first three lines of Table 1.

The various results do not agree.  This discrepancy has inspired some
discussion
\cite{elliskarliner,close}.  Ellis and Karliner \cite{elliskarliner} show, for
example, that the three experiments do agree as long as perturbative QCD
corrections are taken into account.  The results of their analysis, including
these corrections, is shown in the fourth line of Table 1.
The QCD corrections will not exactly apply to the soliton model, but they are
listed to give an idea of the range of values currently under discussion.

\begin{table}[p]
\begin{tabular}{|c||c|c|c|}     \hline
        & $\Delta u$    & $\Delta d$    & $\Delta s$    \\ \hline
EMC     & $.74\pm.06$   & $-.52\pm.06$  & $-.19\pm.06$  \\ \hline
E142    & $.93\pm.03$   & $-.33\pm.03$  & $0.00\pm.04$  \\ \hline
SMC     & $.75\pm.08$   & $-.51\pm.08$  & $-.18\pm.08$  \\ \hline
Ellis \& Karliner
        & $.80\pm.04$   & $-.46\pm.04$  & $-.13\pm.04$  \\ \hline
CQM     & 1.33          & -.33          & 0             \\ \hline
Modified CQM
        & .80           & -.45          & -.20          \\ \hline
\end{tabular}
\caption{}
\end{table}

How does all of this relate to the constituent quark?
In the constituent quark model, the matrix element in Eq.(\ref{delu})
can be related to the helicity of the individual quarks and to the
polarization of the gluons \cite{carlitz} present in the proton:
\[
\Delta u = \frac{4}{3} \int d^{3}r\:<q\uparrow|\bar{\psi}\gamma_{z}\gamma_{5}
\psi|q\uparrow> - \frac{\alpha_{s}}{2\pi}\Delta g
\]
\[
\Delta d = -\frac{1}{3} \int d^{3}r\:<q\uparrow|\bar{\psi}\gamma_{z}\gamma_{5}
\psi|q\uparrow> - \frac{\alpha_{s}}{2\pi}\Delta g
\]
\[
\Delta s = 0 - \frac{\alpha_{s}}{2\pi}\Delta g
\]
As before, $|q\uparrow>$ is a single quark state (of either up or
down flavor), and $\psi$ annihilates that quark.
The gluon contribution $\Delta g$ must be included, because
the current appearing in Eq. (\ref{delu}) can interact via a quark loop with
the gluon sea, which may be polarized.  The prefactors $4/3$ and $-1/3$ come
from the constituent quark model wavefunctions \cite{quigg}.

In the naive constituent quark model
$\int d^{3}r\: <q\uparrow|\bar{\psi}\gamma_{z}\gamma_{5}\psi|q\uparrow>
\equiv s_{q} = 1$ and
$\Delta g = 0$.  This results in line 5 of Table 1, which does not agree well
with experiment.
However, if for some reason
$s_{q}$ turns out to
equal $3/4$, then
$g_{A}=\Delta u - \Delta d = 5/4$, which is very close to the
experimental value.
If in addition $(\alpha_{s}/2\pi)\Delta g = 0.2$, then
the values of $\Delta u,\:\Delta d$, and $\Delta s$ more or less agree
with experiment.  These values are shown in the last line of Table 1.
(The same values
have been used in the context of a relativistic quark model in Ref.
\cite{brodsky}.)  Therefore the spin content of the quark soliton
will be assumed to be
\begin{equation}
s_{q} \equiv \int d^{3}r\:<q\uparrow|j^{3}_{(5)}(r)|q\uparrow> = \frac{3}{4}
\label{j5}
\end{equation}

In the soliton model, the color singlet axial-vector
current $j^{\mu}_{(5)}$ arises only from the Wess--Zumino anomaly term.
In order to compute this current, it is necessary to start from the general
Wess--Zumino Lagrangian which includes both left-- and right--handed fields
\cite{pakrossi}.  Then
\begin{eqnarray}
j^{\mu}_{(5)} & = & \frac{1}{i}
\left( \frac{\partial{\cal L}_{WZ}}{\partial A_{\mu}^{R}}
- \frac{\partial{\cal L}_{WZ}}{\partial A_{\mu}^{L}} \right)_{A^{L}=A^{R}=A}
\nonumber \\
 & = & -\frac{1}{48 \pi^{2}} \epsilon^{\mu \alpha \beta \gamma}
tr[F_{\alpha\beta}( \tilde{R_{\gamma}}-\tilde{L_{\gamma}})]
\label{j5anom}
\end{eqnarray}
and
\begin{eqnarray}
\int d^{3}r\:{\bf j}_{(5)}  & = & \frac{{\bf \Lambda}}{18 \pi^{2} I_{1}}
\int d^{3}r\:\frac{\sin^{2}F}{r} \left[ (1+\gamma)(\omega_{1}^{\prime} +
\omega_{2}^{\prime} + \frac{\omega_{2}}{r}) \right. \nonumber \\
 & &  \left. \mbox{} \:\:\: - \omega_{1}\gamma(1+\gamma)
\frac{1}{r} - \gamma^{\prime}(\omega_{1}+\omega_{2}) \right]
\end{eqnarray}
where again $\bf \Lambda$ is the angular momentum of the soliton.
{}From here, the matrix element required in Eq. (\ref{j5}) can be obtained
easily.

\section{Numerical Results}
\label{results}

The static properties of the quark soliton can now be computed.
It is easiest to look at
dimensionless quantities, since these quantities are determined only by the
three dimensionless parameters $e,\:g,$ and $\nu$.  The fourth parameter $f$
determines the overall scale, and can be fixed later.

The constituent quark can be described by the following two dimensionless
quantities:  the ratio of its color magnetic moment to its magnetic moment,
$\mu_{c}/\beta = 1.12$, and its spin content, $s_{q} = 0.75$.
Requiring that the
soliton's ratio $\mu_{c}/\beta$ take on the physical value of 1.12 will
constrain the permissible values of $e,\:g,\:$ and $\nu$ to lie on a
two--dimensional surface within the three--dimensional parameter space.
Alternatively, requiring that the spin content $s_{q}$ achieve its physical
value of 0.75 will define a different surface within the parameter space.  In
general, these two surfaces will intersect to form a (one--dimensional) curve.
This curve is the family of points where the soliton is a good model of the
constituent quark.  Unfortunately, I have found that these two surfaces do not
intersect.

To begin searching the parameter space for points appropriate to a constituent
quark,
one might first try the point suggested by Ref. \cite{gks} ($e=5.7,\:
\alpha_{s} = 0.28,\: \nu = 0.36$).
However, this gives $\mu_{c}/\beta = 1.94\:$ and $s_{q} = 0.0041\:$ (compared
to the experimental values of 1.12 and 0.75).

The parameter space can be searched using the methods of Refs.
\cite{numericalrecipes} and \cite{acton}, and the results are
summarized in Table 2.  First $\mu_{c}/\beta\:$ is required to equal its
physical value of 1.12, and the input parameters are
varied under the constraint that $\mu_{c}/\beta$ remains constant.  As the
first line of Table 2 shows, $s_{q}$ is less than or
equal to 0.06 under this condition.  The last column of Table 2 gives the
location in parameter space where the maximum value of
$s_{q}$ is achieved.

For the remainder of the parameter space search, $s_{q}$ is fixed at some value
and the input parameters vary under the condition
that $s_{q}$ remains constant.  In each case, $\mu_{c}/\beta$ is bounded from
above; the bounds are listed in Table 2.

Table 2 demonstrates numerically what was said above in words:  the spin
content and the magnetic moments cannot both be simultaneously fitted in this
model.  Any kind of best fit would probably involve making both $s_{q}$ and
$\mu_{c}/\beta$ some fraction (1/4 or less) of their experimental values.
This is the main result of the paper.

\begin{table}[p]
\begin{tabular}{|c|c||c|} \hline
$s_{q}$           & $\mu_{c}/\beta$       & Point where upper limit is reached
\\ \hline \hline
$\leq .06$        & $1.12\:$ (fixed)      & $e = 975,\: \alpha_{s} = 56,\: \nu
= 84$  \\ \hline
$.2\:$ (fixed)    & $\leq .08$            & $e = 970,\: \alpha_{s} = 27,\: \nu
= 88$  \\ \hline
$.4\:$ (fixed)    & $\leq .008$           & $e = 834,\: \alpha_{s} = 20,\: \nu
= 160$ \\ \hline
$.75\:$ (fixed)   & $\leq .0006$          & $e = 808,\: \alpha_{s} = 13,\: \nu
= 232$ \\ \hline
$.75$             & $1.12$                 & Experiment \\ \hline
\end{tabular}
\caption{}
\end{table}

\section{Discussion}
\label{discussion}

The above analysis shows that $\mu_{c}/\beta$ and $s_{q}$ cannot both be of
${\cal O}(1)$ simultaneously.
As discussed in Section \ref{soliton}, however, all of this analysis used
the semi-classical approximation.  This approximation is valid only if the
rotational energy does not contribute much to the total mass; i.e., if
$m_{cl}/M_{TOT} \approx 1$.  However, for all the points listed in Table 2,
$m_{cl}/M_{TOT}$ is between 0.2
and 0.003.  Therefore it is necessary to go beyond the semi-classical
approximation.

In other words, the qualiton model is not a viable model of the constituent
quark if the semi-classical approach is used.  The approach itself is not
valid in the region of parameter space where the model starts to become
interesting. If a better approximation can be used, then what is the hope for
the future of the qualiton model?

The qualiton still faces two obstacles:  its excessive mass, and its strong
interactions.  First, the mass:  within the semi-classical approximation the
product of the mass and the rms radius exceeds a
plausible value, even at its minimum.  While moving in parameter space away
from this minimum in a direction that favors realistic
magnetic moments or spin content, the moments of inertia become so small that
the semi-classical approximation is suspect.
Improving the approximation in the manner suggested after Eq. (\ref{totquant})
may increase the moments of inertia and so lower the
mass,  but this improvement seems unlikely to lower the mass enough to make
$Mr_{rms}$ realistically small.  Another kind of
improvement on the semi-classical approximation, the inclusion of additional
degrees of freedom, will only increase the mass.
So it is likely that no
matter what approximation is used, the mass will be larger than expected for a
consitituent quark.
If the constituent quark
is a soliton, the question is no longer ``What makes the constituent quark so
heavy?'' but ``What makes it so light?''

Second, the strong coupling:  in order to make $s_{q}$ large enough,
$\alpha_{s}$
must be large.  This is because when $\alpha_{s}$ is small ($\alpha_{s}
\leapprox 1$),
$s_{q} \propto \alpha_{s}$.  In the semi-classical approach, the
proportionality
constant is roughly $1.7 \times 10^{-2}$, almost independent of $e$ and $\nu$.
Unless this constant changes by more than two orders of magnitude when the
semi-classical approach is discarded, $\alpha_{s}$ will have to be at least of
${\cal O}(1)$ if $s_{q}$ is to be of ${\cal O}(1)$.
However, any $\alpha_{s} \geapprox 1$
will sabotage the constituent model because the constituents need to be
perturbatively interacting for the model to work.

There is one potential solution to this problem:  the gauge field surrounding
the soliton may screen the particle's charge, so that even when $\alpha_{s}$ is
large, the interactions between qualitons can be treated perturbatively.
Preliminary calculations indicate that some screening does occur.  However
it remains to be seen whether, once the
qualiton is fully quantized (beyond the semi-classical approximation),
this screening is enough to make the qualiton model realistic.

In short, the constituent quark cannot be described by the qualiton in the
semi-classical approximation, and if a better approximation gives the correct
spin and magnetic properties,
the qualiton's large mass and strong interactions will
have to be explained.

Faced with these difficulties, it is tempting to return to the chiral quark
model mentioned in the introduction. One may even wonder
whether constituent quarks exist at all.  Perhaps the constituent
quark model operators and wave functions simply have the right symmetry
properties, and corrections
to their matrix elements are suppressed for some reason (for example, by powers
of $1/N_{c}$ \cite{1/N}).  In any case the success of the constituent quark
model is not yet understood.

\section*{Acknowledgements}

Many thanks to J. D. Jackson and M. Suzuki for their helpful discussions, and
to
S. Johnson, M.-A. Mycek, and A. Papadopoulos for their interest in this
project.
Much of this work was carried out under a fellowship from the Department of
Education
Graduate Assisstance in Areas of National Need program.
This work was also supported in part by the Director, Office of Energy
Research, Office of High Energy and Nuclear Physics, Division of High
Energy Physics of the U.S. Department of Energy under Contract
DE-AC03-76SF00098 and in part by the National Science Foundation under
grant PHY-90-21139.

\pagebreak

{\bf Figure Captions}
\\[1 cm]

{\bf Fig. 1}  The functions $F(\tilde{r})$ and $\gamma(\tilde{r})$ for $1/e =
0,
\: \alpha_{s} = 12.25,\:$ and $\nu=237$.
The soliton has $\mu_{c}/\beta = 2.4\times 10^{-4}$ and $s_{q} = 0.75$.
\\

{\bf Fig. 2}  The functions $\omega_{1},\:\omega_{2},$ and $\omega_{3}$ for the
same input parameters as in Fig. 1.
\\

{\bf Table 1}  Quark contributions to the spin of the proton, using Eqns.
(\protect\ref{bj}) -
(\protect\ref{hyp})
and EMC \protect\cite{emc} (line 1),
E142 \protect\cite{e142} (line 2), or SMC
\protect\cite{smc} data (line 3).  Line 4 gives Ellis and Karliner's analysis
\protect\cite{elliskarliner}.
Line 5 gives the consitiuent quark model (CQM)
prediction, which uses $s_{q} = 1\:$ and $\Delta g = 0$.
Line 6 gives the results of the CQM with the modification that $s_{q} = 3/4\:$
and $(\alpha_{s}/2 \pi) \Delta g = 0.2$.
\\

{\bf Table 2}  The spin and magnetic properties of the soliton.
The first line shows that when the input parameters are varied keeping
$\mu_{c}/\beta$ fixed, $s_{q}$ is always less than the given bound.  The rest
of the table shows
that for $s_{q}\:$ fixed, $\mu_{c}/\beta$ is bounded.  These bounds are
compared with experiment.

\end{document}